# A Knowledge-Grounded Behavioral Reasoning Framework for Training-Free Urban Healthcare OD Prediction

Linzhen Yang[1], Xueliang Liu[1], Chengbo Zhang[2], Meng Meng[1]*

1. South China university of Technology

2. McGill University

**Abstract:** Understanding urban healthcare mobility is essential for healthcare resource allocation, hospital capacity planning, and resilient urban governance. Existing healthcare origin–destination (OD) prediction methods primarily rely on supervised deep learning, requiring expensive model training while providing limited interpretability into the behavioral mechanisms of hospital choice. This paper presents a **training-free urban healthcare OD prediction framework** based on **knowledge-grounded behavioral reasoning**, replacing gradient-based neural network learning with collaborative LLM reasoning over structured urban knowledge. The framework organizes heterogeneous urban information—including healthcare resources, population characteristics, transportation accessibility, weather conditions, and historical mobility—into a unified knowledge graph, enabling evidence retrieval, urban context understanding, behavioral reasoning, preference ranking, and Bayesian verification through a multi-agent reasoning pipeline. Experiments on a real-world Shenzhen dataset containing **137 hospitals** and **6,341 prediction tasks** demonstrate that the proposed framework achieves **CPC = 0.5449**, **Top-1 = 66.19%**, **Top-5 = 68.03%**, and **Top-10 = 74.03%**, consistently outperforming conventional deep learning baselines on all Top-K metrics (e.g., **Top-5 +22.1%** over MLP). Compared with Direct LLM using the same backbone without structured reasoning, the proposed framework improves **CPC by 87.5%** and **Top-1 accuracy by 141%**, demonstrating that **knowledge-grounded behavioral reasoning**, rather than raw LLM capability, is the key to accurate, interpretable, and training-free healthcare OD prediction. These findings highlight the potential of **urban intelligence** to move healthcare mobility modeling beyond data fitting toward knowledge-driven behavioral reasoning.

## 1 Introduction

Healthcare accessibility is a fundamental issue in urban science, public health, and urban computing. Hospital visits flows emerge from the interactions among population distribution, healthcare resources, transportation accessibility, weather conditions, and urban functional areas. These interactions reflect not merely spatial movements but the behavioral processes through which residents make healthcare decisions under heterogeneous urban contexts. Understanding and predicting healthcare origin–destination (OD) flows is therefore important not only for hospital capacity planning, emergency response, healthcare equity, and resilient urban governance, but also for revealing the behavioral mechanisms underlying urban healthcare systems.

Existing spatial interaction models, ranging from the classical gravity model to recent deep learning approaches such as DeepGravity[1], primarily formulate healthcare OD prediction as a data-driven function approximation problem. Although these methods have achieved competitive predictive performance, they exhibit three fundamental limitations. First, they rely on large-scale labeled data and computationally expensive model training. Second, their end-to-end architectures provide limited interpretability, making it difficult to explain why residents choose particular hospitals. More fundamentally, hospital choice is not simply the outcome of statistical feature correlations; rather, it is a knowledge-driven behavioral process jointly shaped by healthcare resources, transportation accessibility, weather conditions, historical mobility patterns, and urban functional characteristics. Such decision-making mechanisms cannot be fully represented through gradient-based feature fitting alone.

Recent advances in Large Language Models (LLMs), knowledge graphs, and intelligent agents provide an opportunity to rethink this problem from the perspective of urban intelligence. Rather than viewing healthcare OD prediction as learning a mapping function, these techniques enable explicit reasoning over heterogeneous urban knowledge and support interpretable decision-making processes. This suggests a new research paradigm in which urban mobility prediction is treated as a behavioral reasoning problem grounded in structured urban knowledge instead of purely a supervised prediction task. It therefore raises a fundamental question: **Can urban healthcare OD prediction be achieved through knowledge-grounded reasoning process rather than conventional neural network training?**

To answer this question, we propose a **Knowledge-Grounded Urban Healthcare Reasoning Framework**, a training-free urban intelligence framework that integrates an urban knowledge graph with collaborative multi-agent LLM reasoning. The framework organizes heterogeneous urban knowledge—including population distribution, healthcare resources, transportation accessibility, weather conditions, and historical mobility—into a unified semantic representation. Through collaborative evidence retrieval, urban context understanding, behavioral reasoning, preference ranking, and Bayesian verification, it explicitly simulates the perception–cognition–decision process underlying hospital choice. Instead of approximating spatial interactions through parameter optimization, the proposed framework generates healthcare mobility predictions by reasoning over urban knowledge, transforming OD prediction from a black-box regression task into an interpretable urban behavioral reasoning process while maintaining competitive predictive performance.

The main contributions of this work are summarized as follows.

**(1) We introduce a knowledge-grounded behavioral reasoning paradigm for urban healthcare OD prediction.** Rather than treating hospital flow prediction as a conventional regression problem, we reformulate it as an urban behavioral reasoning task, demonstrating that structured urban knowledge combined with LLM reasoning can replace conventional gradient-based training while providing transparent behavioral explanations.

**(2) We propose a collaborative multi-agent reasoning framework for modeling urban healthcare decision-making.** By integrating heterogeneous urban knowledge—including healthcare resources, population distribution, transportation accessibility, weather conditions, and historical mobility—into a unified knowledge graph, specialized reasoning agents collaboratively simulate the perception–cognition–decision process that governs hospital choice.

**(3) We develop a Bayesian knowledge-calibrated prediction mechanism that bridges symbolic reasoning and empirical urban behavior.** By combining historical mobility priors with LLM-generated behavioral preferences, the proposed framework produces calibrated probability distributions over all candidate hospitals, improving both prediction reliability and interpretability.

## 2 Related Works

### 2.1 OD Prediction

Healthcare origin–destination (OD) prediction is a fundamental task in urban science, supporting healthcare accessibility assessment, hospital resource allocation, and resilient urban planning. Classical spatial interaction models, such as the Gravity Model, describe healthcare flows through attraction and distance-decay mechanisms, while DeepGravity [错误!未定义书签。] 、replaces handcrafted interaction functions with deep neural networks to improve predictive performance. More recently, Transformer-based models [2] have further enhanced urban mobility prediction by modeling complex spatial dependencies. Despite these advances, existing approaches predominantly formulate healthcare mobility as a supervised function approximation problem, learning statistical mappings from features to flows. Consequently, they provide limited understanding of the behavioral mechanisms underlying hospital choice and require computationally intensive model training.

### 2.2 Urban Knowledge Representation

Understanding healthcare mobility requires integrating heterogeneous urban information, including population distribution, healthcare resources, transportation accessibility, weather conditions, and historical mobility. In flow prediction, Knowledge graphs provide a natural representation for organizing these heterogeneous urban entities and their relationships into structured semantic networks[3]. Existing studies have investigated knowledge graph construction and reasoning frameworks, while recent works such as KG-Thought demonstrate that combining knowledge graphs with LLMs improves knowledge-intensive reasoning tasks[4]. However, most existing studies employ knowledge graphs primarily as information repositories or retrieval tools. They rarely exploit structured urban knowledge to explicitly model the behavioral reasoning process that drives hospital choice[5].

### 2.3 LLM Reasoning and Multi-Agent Systems

Recent advances in Large Language Models have shifted artificial intelligence from statistical prediction toward explicit reasoning over structured knowledge. Chain-of-Thought and ReAct [6] demonstrate that decomposing complex decision-making into interpretable reasoning steps substantially improves reasoning quality, while MetaGPT[7] and Generative Agents[8] extend this capability to collaborative multi-agent systems. These studies suggest that complex decision processes can be modeled through collaborative reasoning rather than solely through parameter optimization. Nevertheless, existing LLM-based urban intelligence frameworks mainly focus on autonomous task execution or general urban question answering, instead of reasoning about individual mobility decisions grounded in heterogeneous urban knowledge.

### 2.4 From Reasoning to Reliable Prediction

Although LLMs provide strong reasoning capability, their outputs remain probabilistic and may deviate from empirical behavioral patterns observed in real cities. Therefore, prediction reliability requires connecting symbolic reasoning with observed urban behavior. Existing calibration methods have been widely adopted to improve prediction confidence in deep learning models[9]. but relatively few studies investigate how reasoning outputs can be calibrated using historical urban mobility evidence[9]. This motivates integrating Bayesian calibration with LLM-generated behavioral preferences, allowing knowledge-driven reasoning to remain consistent with empirical healthcare mobility patterns.

### 2.5 Summary and Positioning

Table 1 summarizes representative approaches from the perspectives of prediction paradigm, knowledge utilization, reasoning capability, and interpretability. Existing deep learning models, such as MLP and DeepGravity, achieve competitive prediction performance but rely on supervised training and treat healthcare mobility primarily as a regression problem. Knowledge graph approaches provide structured urban knowledge but generally lack explicit behavioral reasoning. Recent LLM-based urban intelligence frameworks introduce reasoning capability, yet seldom integrate heterogeneous urban knowledge with probabilistic calibration for urban mobility prediction. Furthermore, most existing approaches either optimize predictive models or generate reasoning independently, without explicitly bridging behavioral reasoning and empirical urban observations. To provide a fair comparison, we additionally include Direct LLM as a lightweight baseline using the same Qwen model without knowledge graph augmentation, collaborative reasoning, or Bayesian calibration. In contrast, the proposed framework reformulates healthcare OD prediction as a **knowledge-grounded urban behavioral reasoning problem**, integrating heterogeneous urban knowledge, collaborative multi-agent reasoning, and Bayesian calibration within a unified training-free urban intelligence framework.

**Table 1. Comparison of existing methods for OD prediction and LLM-based reasoning.**

| Method | Prediction Paradigm | Training | Urban Knowledge | Behavioral Reasoning | Calibration | Interpretability |
|---|---|---|---|---|---|---|
| Gravity | Analytical | No | ✗ | ✗ | ✗ | Low |
| DeepGravity | Regression | Supervised | ✗ | ✗ | ✗ | Low |
| MLP | Regression | Supervised | ✗ | ✗ | ✗ | Low |
| Direct LLM | Reasoning | None | ✗ | ✓ | ✗ | Medium |
| **Ours** | **Behavioral Reasoning** | **Training-free** | ✓ | ✓ | ✓ | **High** |

## 3 Method

### 3.1 Overview

Given a date $t$ and an origin grid $o$, the objective is to predict the conditional probability distribution over candidate hospitals: $P(H \mid o, t)$, which represents the probability that residents from origin $\boldsymbol{o}$ choose hospital $h$ on date $t$.

Unlike conventional OD prediction methods that directly learn a mapping from input features to flow distributions through gradient optimization, we formulate healthcare mobility prediction as a **knowledge-grounded urban reasoning problem**. Rather than approximating hospital choice by statistical feature fitting, the proposed framework explicitly infers residents' decision-making process by reasoning over heterogeneous urban knowledge, including healthcare resources, transportation accessibility, weather conditions, population characteristics, and historical mobility patterns. The final probability distribution is then obtained by integrating reasoning outcomes with empirical behavioral evidence.

**Problem Statement.** Let $\boldsymbol{O} = \{\boldsymbol{o}_1, \boldsymbol{o}_2, ..., \boldsymbol{o}_n\}$ denote the set of residential origin grids and $\boldsymbol{H} = \{\boldsymbol{h}_1, \boldsymbol{h}_2, ..., \boldsymbol{h}_m\}$ the set the candidate hospitals in a city. Given an origin-date pair $(o, t)$, the task is to estimate $P(H \mid o,t) = [P(h_1 \mid o,t), \cdots, P(h_M \mid o,t)]$, subject to $\sum_{j=1}^{M} P(h_j \mid o,t) = 1$

Instead of directly predicting $P(h_j|o,t)$, we explicitly introduce an intermediate **behavioral state** that characterizes the urban decision-making context,

$$B(o,t) = \{N, A, C, P\} \tag{1}$$

where $N$ denotes inferred healthcare demand, $A$ represents accessibility, $C$ describes healthcare service capability, $P$ denotes behavioral preference. The prediction process is therefore formulated as:

$$G \rightarrow B(o,t) \rightarrow P(H \mid o,t) \tag{2}$$

where $G$ is the urban knowledge graph. This formulation separates **knowledge representation**, **behavioral reasoning**, and **probability estimation**, transforming healthcare OD prediction from a black-box regression problem into an interpretable urban behavioral reasoning process.

### Computational Framework:

Following this formulation, the proposed framework consists of three computational stages:

$$P(H \mid o,t) = f_{Cal}!\left(f_{Reason}!\left(f_{KG}(G,o,t)\right)\right) \tag{3}$$

**Urban Knowledge Representation** constructs a structured urban context by retrieving heterogeneous information from the knowledge graph and organizing it into a unified behavioral representation.

**Behavioral Preference Reasoning** infers residents' hospital preferences through explicit multi-step reasoning over the urban context, simulating the perception–cognition–decision process underlying healthcare choice.

Behavioral Calibration integrates reasoning-derived preferences with historical behavioral priors through Bayesian calibration to generate a probability distribution over all candidate hospitals while maintaining consistency with observed urban mobility patterns.

These three computational stages are implemented through five collaborative functional modules—Retrieval, Analysis, Coordinator, Ranking, and Verification—which together constitute the proposed training-free reasoning framework. Figure 1 illustrates the overall computational architecture:

$$G \rightarrow C(o,t) \rightarrow B(o,t) \rightarrow P(H \mid o,t) \tag{4}$$

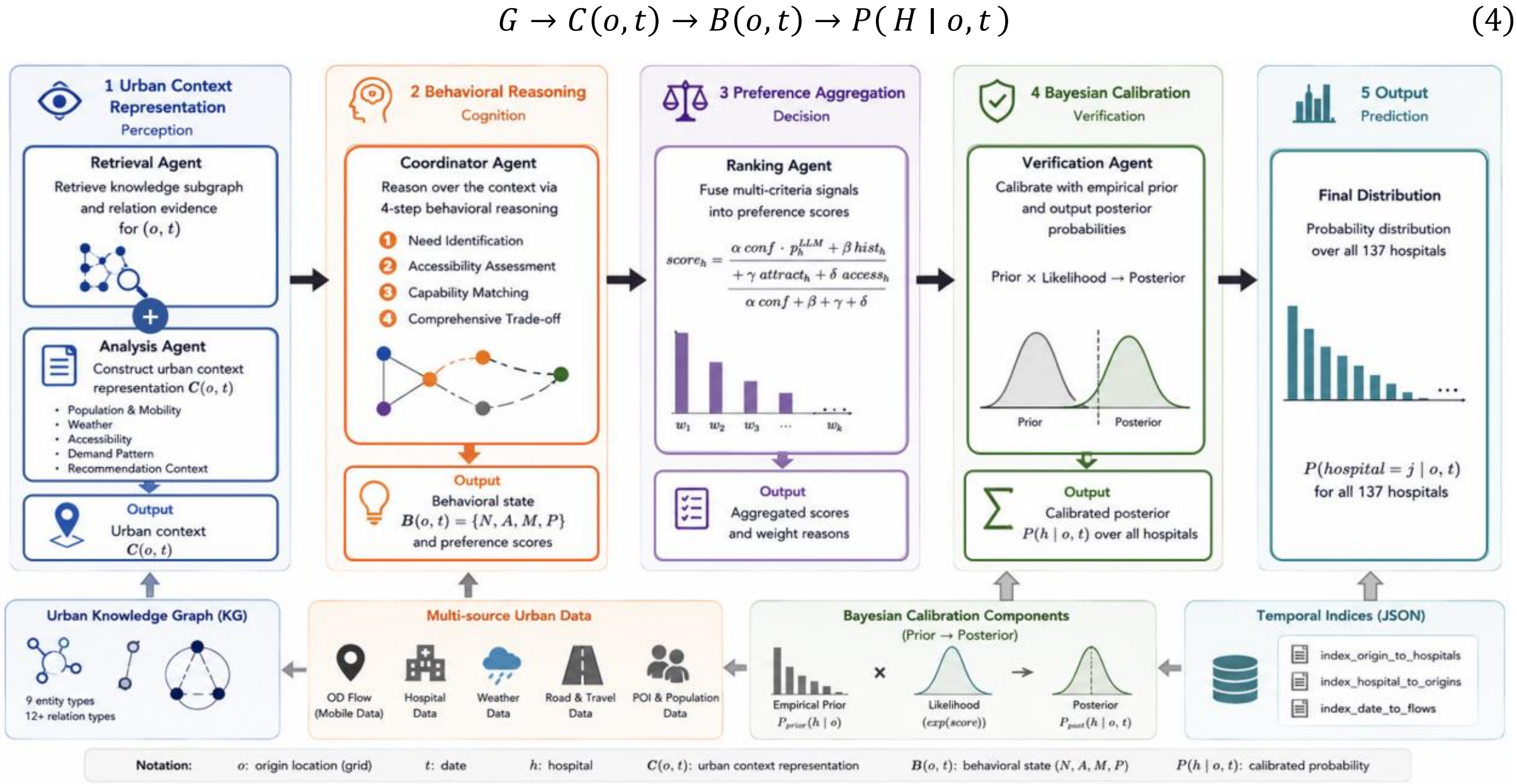

**Fig. 1. Five-module reasoning chain driven by LLM (Qwen3.6-35B-A3B).**

### 3.2 Knowledge Graph Construction

To support knowledge-grounded behavioral reasoning, heterogeneous urban information is organized into a Behavioral Urban Knowledge Graph, which provides structured evidence for subsequent reasoning (Fig. 2). The graph integrates four urban data sources, including healthcare OD flows, hospital information, weather observations, and road networks.

The Knowledge Graph contains nine entity types (Origin, Hospital, Road, Weather, Date, Population Pattern, Hospital Specialty, Travel Condition, and Choice Context) and twelve semantic relation types describing behavioral similarity, healthcare capability, accessibility, temporal context, and environmental influence. These relations enable multi-hop evidence retrieval for healthcare destination reasoning instead of treating urban features as independent variables.

To support efficient inference, three inverted indices (origin-to-hospital, hospital-to-origin, and date-to-flow) are constructed for fast retrieval of behaviorally relevant subgraphs during reasoning.

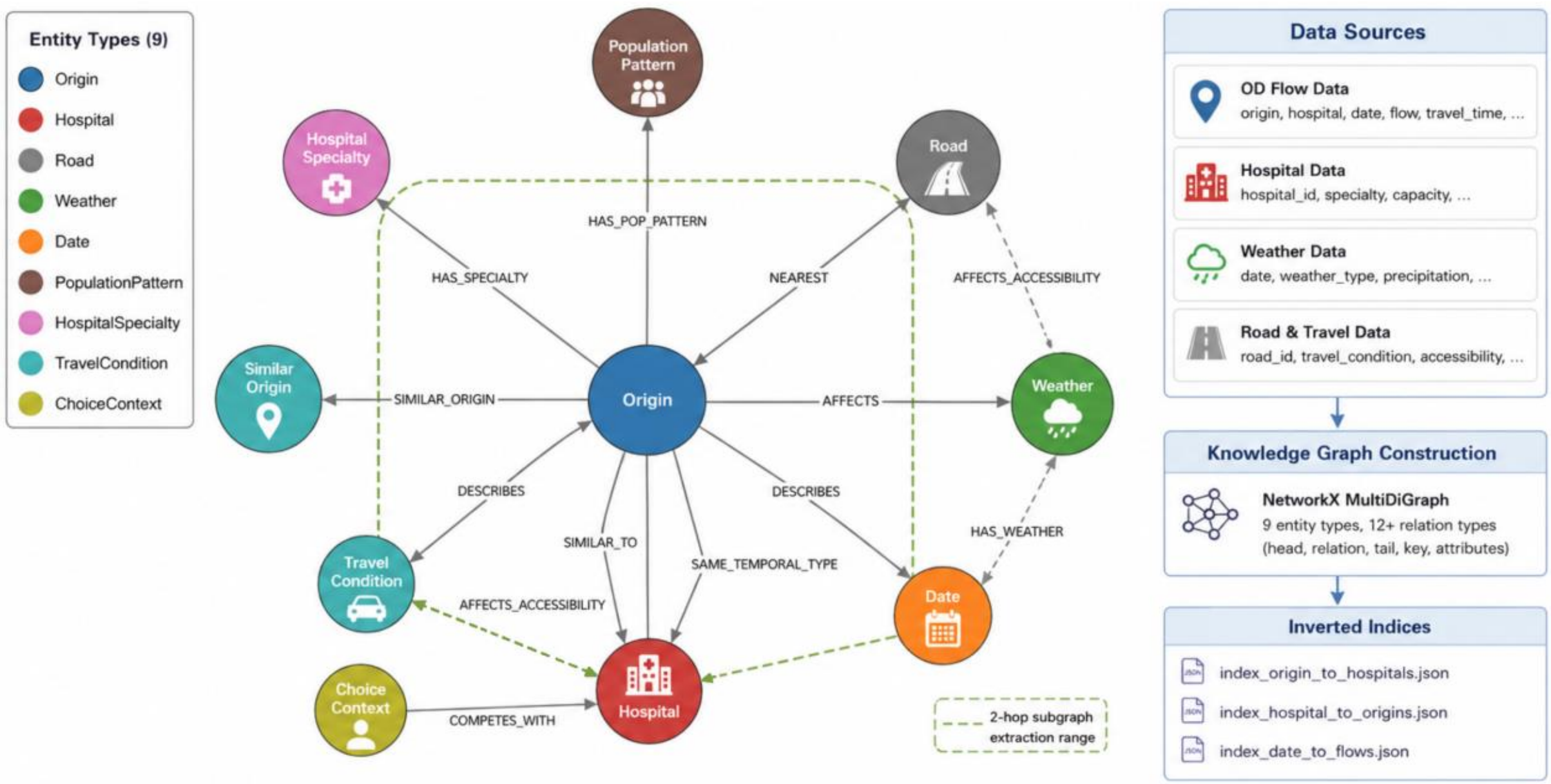

**Fig. 2. Behavioral Urban Knowledge Graph, representing heterogeneous urban knowledge through nine entity types and twelve semantic relations to support healthcare behavioral reasoning.**

### 3.3 Urban Context Representation

Following the problem formulation in Section 3.1, the first stage aims to construct an **urban context representation** for each origin-date pair

Four relation evidence path types are extracted:

- **Behavior transfer:** Origin → SIMILAR_ORIGIN → Origin → visited → Hospital
- **Hospital competition:** Hospital → COMPETES_WITH → Hospital
- **Capability similarity:** Hospital → SIMILAR_TO → Hospital
- **Environmental influence:** Date → Weather → Road → Accessibility → Hospital

Instead of using these relations directly, the framework summarizes them into a structured urban behavioral context**:**

$$C(o,t) = \{C_{pop}, C_{mob}, C_{weather}, C_{access}, C_{demand}, C_{recommend}\} \tag{5}$$

which describes six complementary aspects of the decision environment, including population characteristics, mobility patterns, weather conditions, accessibility, healthcare demand, and recommendation context.

The next step compute rule-based 6-dimensional scene features that summarize the prediction context. The Analysis Agent consumes the RetrievalResult and assembles a complete Scene object with six components: Population Scene, Mobility Scene, Weather Scene, Accessibility, Demand Pattern, and Recommendation. Among these components, accessibility is quantified as

$$A = e^{-d/5} \times max(0.5, 1 - 0.05 \cdot Rain) \times min(1.2, 1 + 0.1 \cdot RD) \tag{6}$$

where $d$ is the distance in km, $Rain$ is the rainfall (mm/day), and RD is the road density (m/km2). Healthcare demand is further inferred from urban functional composition using rule-based semantic knowledge. Residential-dominant regions are associated with routine healthcare demand, work-dominant regions with emergency-oriented demand, and leisure-oriented regions with specialist demand. The resulting urban context representation provides the structured evidence for subsequent behavioral reasoning.

### 3.4 Behavioral Preference Reasoning

The second stage infers residents' hospital preferences by explicitly modeling the behavioral reasoning process underlying healthcare decision-making. Given the urban context representation $\boldsymbol{C}(o, t)$, the framework constructs a structured reasoning prompt that combines six complementary information sources:

- **origin and temporal information,**
- **knowledge graph evidence,**
- **historical reasoning memory,**
- **urban context representation,**
- **healthcare domain knowledge,**
- **candidate hospital descriptions**

Instead of directly predicting hospital probabilities, the framework formulates hospital selection as a four-stage behavioral reasoning process,

$$C(o,t) \longrightarrow B(o,t) \tag{7}$$

where the behavioral state:

$$B(o,t) = \{N, A, M, P\} \tag{8}$$

contains inferred healthcare Need, Accessibility, Capability Matching, and Behavioral Preference.

The reasoning process sequentially evaluates

- **healthcare demand identification,**
- **accessibility assessment,**
- **hospital capability matching,**
- **multi-factor behavioral trade-off.**

Each reasoning stage explicitly explains how heterogeneous urban knowledge contributes to hospital selection, producing both preference scores and natural-language reasoning traces.

To improve reasoning consistency across recurring urban scenes, the framework further maintains a Scene Pattern Memory, indexed by:

$$\left(poi_{dominant}, temporal_{type}, weather_{level}\right) \tag{9}$$

whose influence gradually decays over time. Similar urban contexts therefore reuse previous behavioral experience while remaining adaptive to evolving environmental conditions.

This behavioral reasoning stage is implemented through the Coordinator Agent.

### 3.5 Preference Aggregation

The behavioral reasoning stage produces qualitative hospital preferences, which are subsequently transformed into quantitative ranking scores through a preference aggregation model. Rather than relying solely on LLM outputs, the framework integrates multiple complementary behavioral signals, including LLM reasoning confidence, historical hospital preference, hospital attractiveness, and accessibility. The final preference score is computed as:

$$score_h = \frac{\alpha conf \cdot p_h^{LLM} + \beta hist_h + \gamma attract_h + \delta access_h}{\alpha conf + \beta + \gamma + \delta} \tag{10}$$

where the default weights are $\alpha$ = 0.4, $\beta$ = 0.3, $\gamma$ = 0.2, $\delta$ = 0.1.

Since different urban scenarios emphasize different decision factors, the aggregation weights are dynamically adjusted whenever the reasoning preference conflicts with historical behavioral patterns. In such cases, the LLM performs an additional arbitration process that re-estimates the contribution of each behavioral factor while simultaneously generating interpretable explanations for every weight adjustment.

Consequently, the final preference score reflects both explicit behavioral reasoning and empirical urban knowledge rather than either source individually.

### 3.6 Step 4: Bayesian Behavioral Calibration

Although behavioral reasoning produces reliable preference rankings, the resulting scores remain local estimates over the retrieved candidate hospitals. The final stage therefore transforms these reasoning outputs into a calibrated probability distribution over all hospitals. The calibration process combines historical behavioral priors with reasoning-derived likelihoods through Bayesian inference. Historical prior is computed as:

$$P_{prior}(h \mid o) = \frac{flow_{o,h}}{\sum_h flow_{o,h}} \tag{11}$$

The likelihood generated from behavioral reasoning is:

$$L_h = (p_h^{LLM})^{\tau} \cdot attract_h^{\eta} \cdot e^{-d/R_0} \tag{12}$$

Posterior preference is then estimated by

$$Posterior_h = P_{prior}(h \mid o)^{w} \cdot L_h \tag{13}$$

followed by Softmax normalization

$$P(h \mid o, t) = \frac{\exp(\log s_h / T)}{\sum_{h'} \exp(\log s_{h'} / T)} \tag{14}$$

Here, $\tau$ controls reasoning confidence, $\eta$ regulates hospital attractiveness, $R_0$ determines spatial decay, and $w$ balances historical evidence with behavioral reasoning.

To ensure prediction reliability, three consistency constraints are further imposed:

- **probability consistency,**
- **ranking consistency,**
- **historical distribution consistency.**

If any constraint is violated, the framework increases the contribution of historical evidence and repeats Bayesian updating until convergence or a predefined iteration limit is reached.

An important design principle of the proposed framework is the separation between behavioral reasoning and probability propagation. Behavioral reasoning is performed only on the retrieved candidate hospitals to reduce LLM inference cost, whereas Bayesian calibration propagates these preferences to the complete hospital set, producing a calibrated probability distribution over all candidate hospitals while maintaining computational

# 4 Experiment Design

## 4.1 Dataset

We evaluate the proposed framework using a real-world urban healthcare mobility dataset collected in Shenzhen. The dataset combines anonymized mobile phone signaling records with heterogeneous urban knowledge, including healthcare resources, population distribution, urban functional areas, transportation accessibility, and meteorological observations. Hospital visit trajectories are extracted from behavior chains to construct daily healthcare origin–destination (OD) networks, providing realistic observations of urban healthcare mobility.

The study area contains 137 geocoded hospitals and approximately 10,000 grids at a spatial resolution of 1 km. Population density is obtained from LandScan, POI composition is categorized into residential, work, leisure, and other functions, daily rainfall and temperature are interpolated from meteorological stations using Kriging, while hospital level, bed capacity, and road density are collected from official municipal datasets. These heterogeneous datasets jointly constitute the urban knowledge graph described in Section 3.

Following a temporal evaluation protocol, the first 25 days (September 1–25, 2025) are used to construct historical behavioral knowledge, while the remaining five days (September 26–30, 2025) are reserved exclusively for testing. This setting yields 6,341 healthcare mobility prediction tasks without introducing information leakage across time.

## 4.2 Baselines

To evaluate the effectiveness of the proposed knowledge-grounded behavioral reasoning paradigm, we compare our framework with representative approaches from three prediction paradigms.

The first category consists of analytical and data-driven prediction models, including: (1) Gravity, the classical spatial interaction model with the distance-decay exponent optimized by grid search; (2) MLP, a three-layer multilayer perceptron that directly learns OD probabilities from heterogeneous urban features; and (3) DeepGravity, a ten-layer deep neural network representing state-of-the-art supervised spatial interaction modeling. Following a unified temporal evaluation protocol, all supervised baselines are trained on the first 25 days (September 1–25, 2025) and evaluated on the same five-day test period (September 26–30, 2025), ensuring a fair comparison under identical training and testing conditions.

The second category represents LLM-based reasoning without structured urban knowledge. We adopt Direct LLM using the same Qwen3.6-35B-A3B model as our framework, but with a minimal prompt that excludes the urban knowledge graph, urban context representation, scene pattern memory, collaborative reasoning, and Bayesian calibration. The generated preference scores are directly converted into probability distributions through Softmax normalization. This baseline isolates the contribution of the proposed knowledge-grounded behavioral reasoning framework from the reasoning capability of the LLM itself.

### 4.3 Metrics

Evaluation metrics include: $CPC$ (Common Part of Commuters, Sorensen-Dice coefficient) measuring overall flow agreement; $NRMSE$ (Normalized Root Mean Squared Error, std-normalized) measuring regression accuracy; $JSD$ (Jensen-Shannon Divergence) measuring probability distribution shape difference; and Top-5/Top-10 hit rates measuring ranking quality. For hospital flow prediction, $Top-K$ hit rates directly reflect the model's ability to identify the correct hospital and carry stronger practical significance. $NRMSE$ is reported using std-normalization to align with the DeepGravity evaluation protocol. The formal definitions are as follows.

$$CPC = 2 \times \frac{\Sigma min(p_i, g_i)}{\Sigma p_i + \Sigma g_i} \tag{15}$$

$$NRMSE = \frac{\sqrt{(\Sigma(\hat{y}_i - y_i)^2/N)}}{std(y)} \tag{16}$$

$$Top-k = \frac{\Sigma[rank(h_{true}) \leq k]}{N} \tag{17}$$

$$JSD = \frac{1}{2} KL(P||M) + \frac{1}{2} KL(Q||M), M = \frac{P + Q}{2} \tag{18}$$

where $p_i$ and $g_i$ are predicted and ground-truth flows, $\hat{y}_i$ and $y_i$ are predicted and true flow values, $h_{true}$ is the actual target hospital, and $P$, $Q$ are the two distributions being compared with $M$ their midpoint distribution.

### 4.4 Parameter Sensitivity Analysis

Grid Search is performed over the Bayesian calibration hyperparameters $(\tau, \eta, R^0, T)$. The search ranges are: τ in $\{0.5, 1.0, 1.5, 2.0\}$, η in $\{0.3, 0.5, 0.7, 1.0\}$, $R^0$ in {3.0, 5.0, 7.0, 10.0}, and $T$ in $\{0.5, 1.0, 1.5\}$. Fig. 3 visualizes the impact of each hyperparameter on CPC and Top-1 while keeping the remaining parameters at their optimal values. As shown in panel (a), the LLM temperature τ exhibits a clear peak at $\tau$ =1.0: lower values ($\tau$ =0.5) flatten the score differences between hospitals, reducing ranking discrimination (CPC=0.5423), while higher values ($\tau$=1.5, 2.0) over-amplify small variations and introduce instability (CPC=0.5365 at τ=2.0). Panel (b) shows that the attractiveness exponent η has a moderate optimum at $\eta$=0.5; lower η under-weights hospital capacity and level, whereas higher η over-emphasizes these attributes at the expense of distance signals. Panel (c) indicates that the distance decay radius $R^0$=5.0 km best balances local accessibility and regional hospital choice, with overly short radii ($R^0$=3.0) missing viable candidates and overly long radii ($R^0$=10.0) diluting the distance signal. Panel (d) demonstrates that the softmax temperature $T$=1.0 provides the best probability calibration, as T=0.5 produces over-confident distributions (CPC=0.5428) and $T$ =1.5 over-smooths them (CPC=0.5435). The optimal configuration $\tau$=1.0, $\eta$=0.5, $R^0$=5.0, $T$=1.0 is used for all reported results.

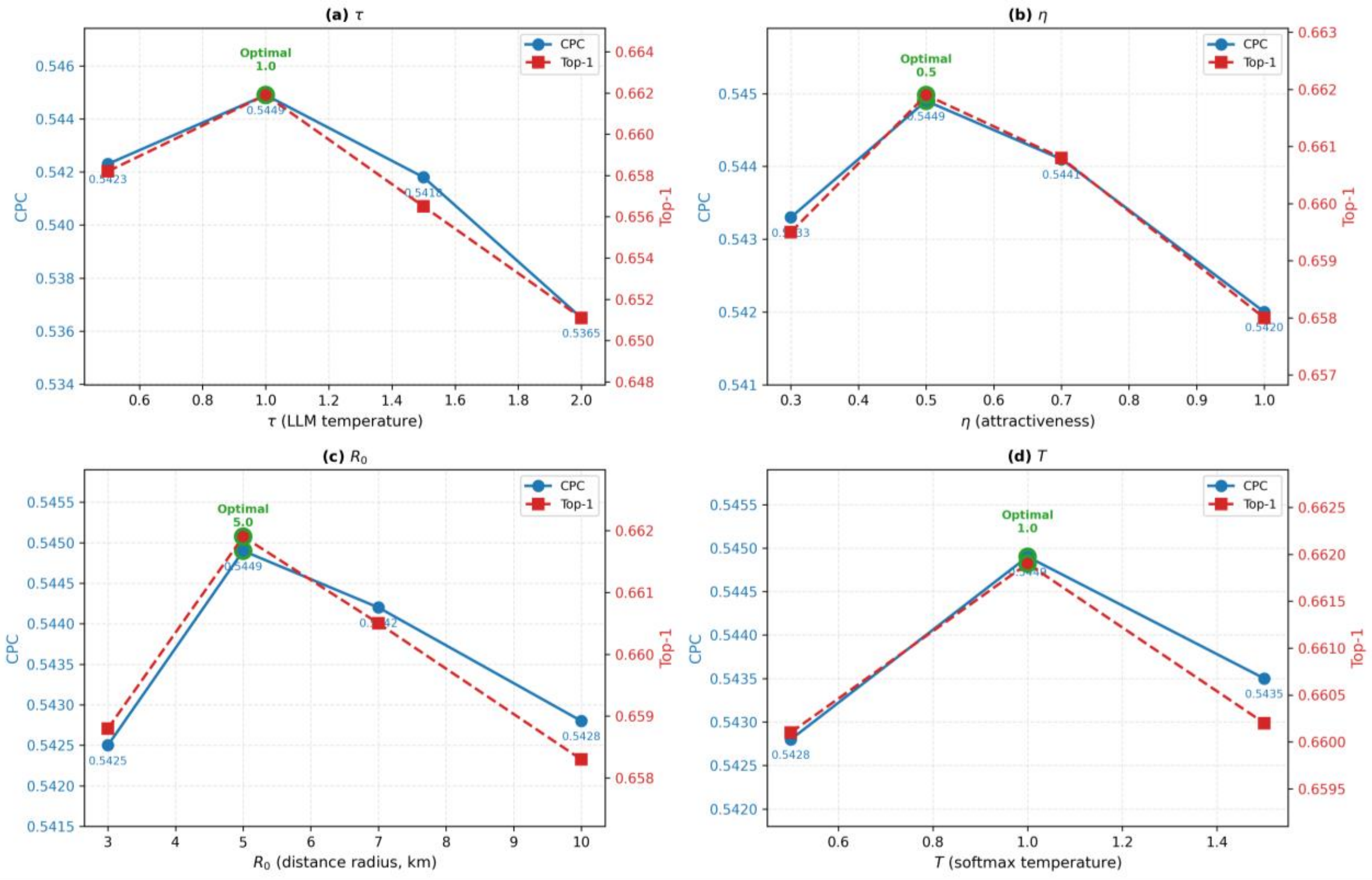

Fig. 3. Parameter Sensitivity Analysis results.

# 5 Experiment Results

## 5.1 Overall Performance

Fig. 4 summarizes the quantitative comparison of all methods on the five-day real-world test set (September 26–30, 2025). Overall, the proposed knowledge-grounded behavioral reasoning framework achieves competitive overall prediction accuracy while substantially outperforming all baselines in hospital ranking performance.

Specifically, the proposed framework achieves a CPC of 0.5449, approaching the supervised deep learning models (DeepGravity: 0.6002; MLP: 0.5850), while obtaining an NRMSE of 0.7318, comparable to DeepGravity (0.6973) and superior to MLP (0.7513). More importantly, it consistently achieves the highest Top-1 (0.6619), Top-5 (0.6803), and Top-10 (0.7403) accuracies, improving Top-5 and Top-10 by 22.1% and 42.6%, respectively, over the strongest baseline. These improvements indicate that explicit behavioral reasoning is particularly effective for identifying the most plausible hospital destination, which is the primary objective of healthcare OD prediction.

To further isolate the contribution of the proposed framework, we compare it with Direct LLM, which employs the identical Qwen model but without urban knowledge graphs, collaborative reasoning, or Bayesian calibration. Direct LLM achieves only 0.2908 CPC and 0.2742 Top-1, whereas the proposed framework improves these metrics by 87.5% and 141%, respectively. This substantial performance gap demonstrates that prediction quality originates primarily from knowledge-grounded behavioral reasoning, rather than the reasoning capability of the LLM alone.

Fig. 5 further illustrates the Top-K hit-rate curves. The proposed framework consistently dominates across all values of KKK, with the largest advantage observed at Top-5 and Top-10. In contrast, DeepGravity and MLP exhibit slower improvements as KKK increases, indicating that their predicted probabilities are more diffusely distributed across candidate hospitals. These results suggest that behavioral reasoning generates more discriminative hospital preference rankings than conventional feature-based regression models.

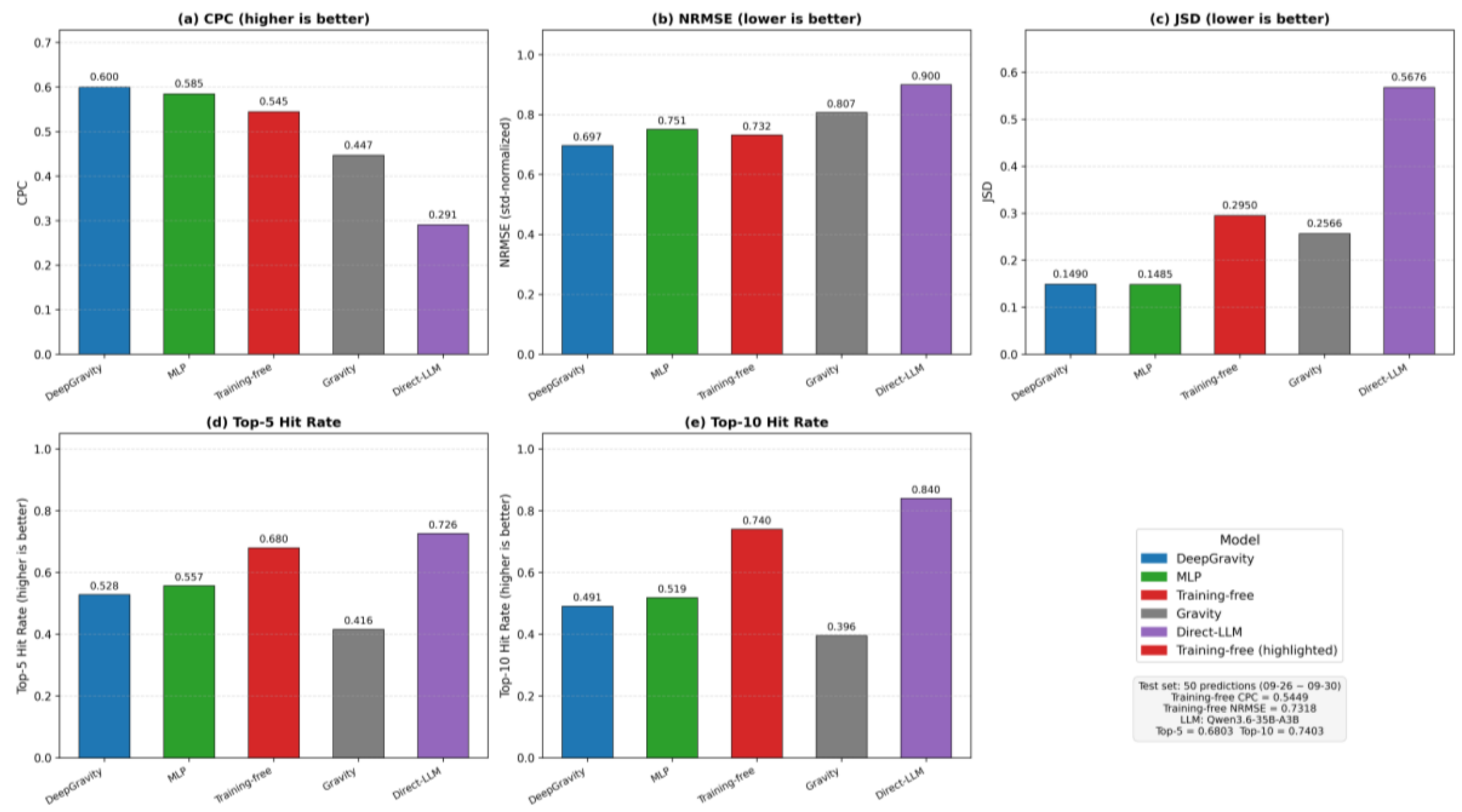

**Fig. 4. Overall comparison validating the effectiveness of behavioral reasoning.**

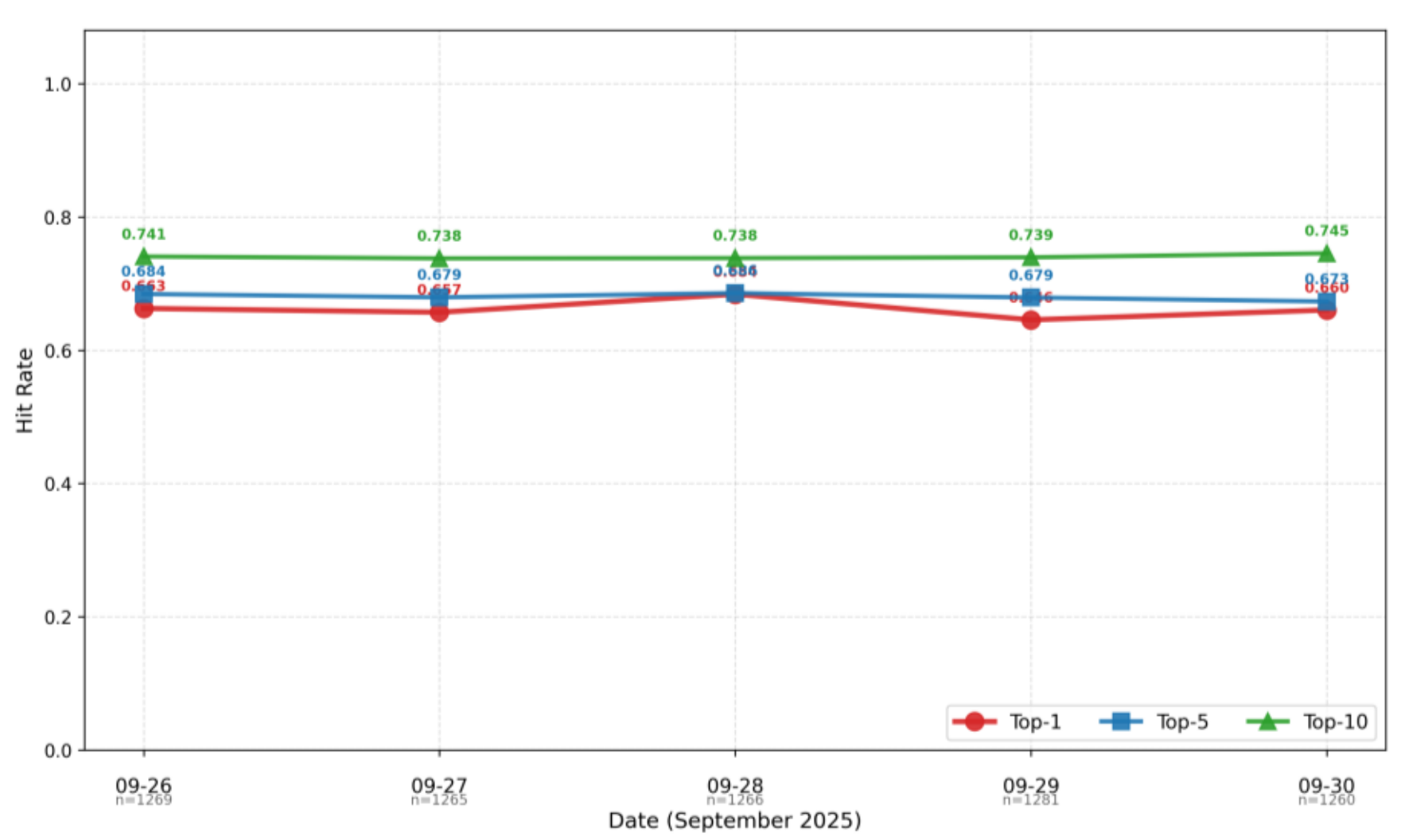

**Fig. 5. Top-K ranking performance.**

## 5.2 Geographic Distribution

Fig. 6 compares the spatial distribution of predicted hospital inflows with the observed healthcare mobility patterns. Beyond overall prediction accuracy, this experiment evaluates whether different prediction paradigms can reproduce the geographic organization of urban healthcare demand.

The observed hospital inflows exhibit a clear spatial concentration pattern, in which a limited number of high-capacity hospitals attract a disproportionately large share of visits while numerous hospitals receive relatively low demand. This pattern reflects the combined influence of healthcare service capacity, transportation accessibility, and distance-decay effects.

Compared with analytical and supervised learning baselines, the proposed framework more accurately reproduces this spatial organization. Gravity excessively concentrates flows toward the nearest hospitals, whereas DeepGravity and MLP tend to over-disperse demand across many destinations. Direct LLM exhibits a different bias by concentrating flows toward frequently mentioned hospitals without adequately considering accessibility or service capability, leading to substantially lower overall accuracy.

In contrast, the proposed framework effectively balances these competing factors through knowledge-grounded behavioral reasoning. By explicitly integrating healthcare resources, accessibility, weather conditions, and historical mobility into the reasoning process, it successfully captures both the concentration around major hospitals and the long-tail distribution of smaller facilities, producing a spatial pattern that most closely matches the observed urban healthcare system.

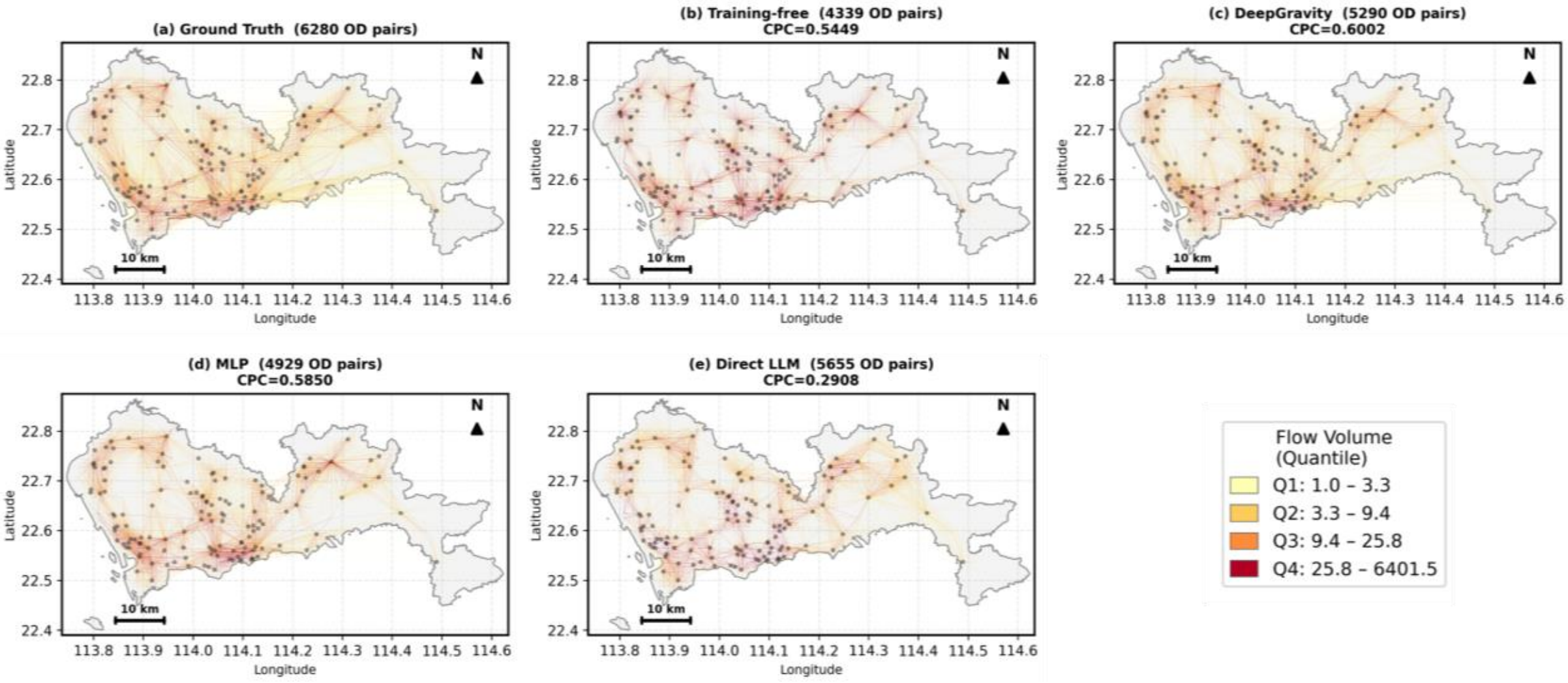

**Fig. 6. Recovery of urban healthcare spatial organization.**

### 5.3 Prediction Accuracy

Fig. 7 compares predicted and observed hospital flows to evaluate whether the proposed framework produces well-calibrated probability distributions rather than accurate rankings alone.

Compared with supervised learning models, the proposed framework generates prediction distributions that more closely resemble empirical healthcare mobility patterns. A small number of major hospitals receive relatively high probabilities, while long-tail hospitals retain non-zero probabilities after Bayesian calibration, ensuring complete coverage of all 137 hospitals without collapsing toward only a few dominant destinations.

DeepGravity and MLP produce more dispersed probability distributions, reflecting their regression-oriented optimization objective. Direct LLM, on the other hand, overemphasizes high-frequency hospitals because its reasoning lacks historical behavioral constraints. The proposed framework effectively bridges these two extremes through Bayesian behavioral calibration, which combines reasoning-derived preferences with empirical mobility priors.

Consequently, the predicted flows exhibit stronger agreement with the observed distributions, particularly for high-demand hospitals where behavioral reasoning provides more reliable destination ranking. These results confirm that Bayesian calibration successfully anchors LLM reasoning to real urban healthcare behaviors, improving both prediction reliability and distributional consistency.

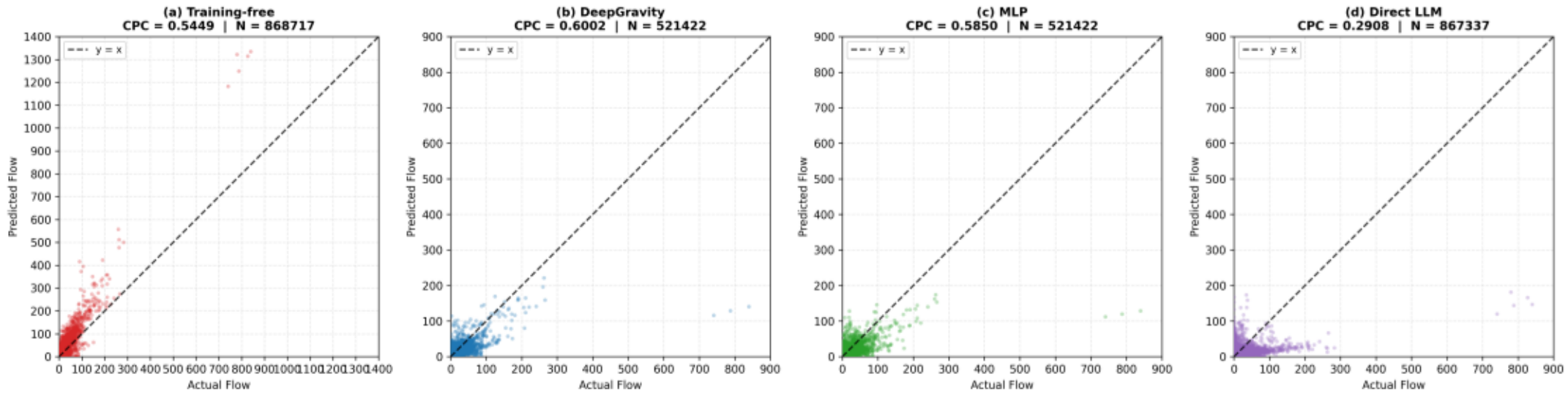

**Fig. 7. Probability calibration.**

### 5.4 Interpretability

A key advantage of the proposed framework is that every prediction is accompanied by an explicit behavioral reasoning process rather than only a final probability estimate. Figure 8 (Appendix A) presents three representative healthcare mobility scenarios illustrating how the framework performs urban behavioral reasoning.

Across all cases, hospital selection is explained through four complementary reasoning levels: (1) inferred healthcare demand and confidence, (2) accessibility–capability–cost trade-offs, (3) supporting evidence retrieved from the urban knowledge graph, and (4) dynamic adjustment of behavioral preferences through collaborative reasoning and Bayesian calibration. Unlike conventional deep learning models, every modification of the predicted ranking is explicitly justified by both reasoning traces and supporting urban knowledge.

These examples demonstrate that the proposed framework provides transparent explanations for hospital choice while preserving competitive predictive performance. Such interpretability is particularly valuable for healthcare applications, where understanding **why** a hospital is recommended is often as important as accurately predicting **which** hospital will be selected.

## 6 Ablation Study

To understand how different behavioral reasoning mechanisms contribute to the proposed framework, we conduct a series of ablation experiments by progressively removing one mechanism while keeping all other components unchanged. Figure 8 summarizes the results across all evaluation metrics. Rather than validating individual software modules, these experiments examine the contribution of five complementary reasoning mechanisms, including structured urban knowledge, urban context representation, behavioral memory, preference arbitration, and Bayesian behavioral calibration.

Overall, removing any mechanism consistently degrades prediction performance, demonstrating that the effectiveness of the proposed framework arises from the joint interaction of multiple behavioral reasoning mechanisms rather than from a single component. Among them, knowledge graph reasoning and Bayesian calibration contribute the largest performance gains, whereas scene understanding, behavioral memory, and dynamic arbitration further improve reasoning stability, consistency, and robustness across diverse urban scenarios.

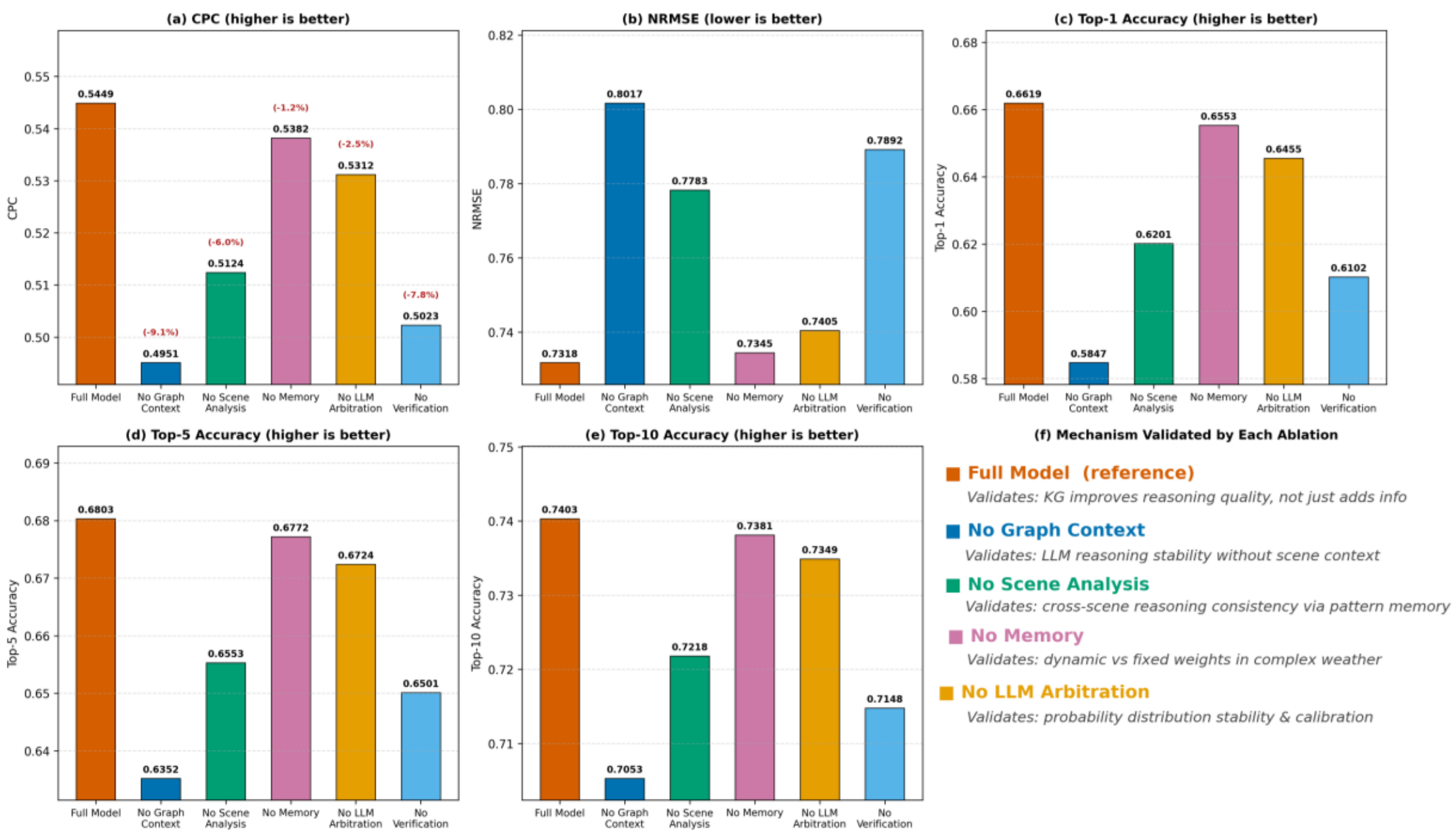

**Fig. 8. Behavioral mechanism validation.**

**No Calibration:** Removing Bayesian behavioral calibration results in the largest overall performance degradation, with CPC decreasing from **0.5449** to **0.5023** (−7.8%), Top-1 decreasing from **0.6619** to **0.6102**, and NRMSE increasing from **0.7318** to **0.7892**. These results indicate that behavioral reasoning alone is insufficient to produce reliable probability distributions. Bayesian calibration plays a critical role by anchoring LLM-derived preferences to empirical healthcare mobility patterns, thereby improving both ranking accuracy and distributional consistency.

**No KG Context:** Removing the urban knowledge graph causes CPC to decrease from **0.5449** to **0.4951** (−9.1%) and Top-1 from **0.6619** to **0.5847** (−11.7%). Without structured urban knowledge, the reasoning process can no longer exploit cross-origin behavioral transfer, hospital competition, or capability similarity encoded in the knowledge graph. These findings demonstrate that structured urban knowledge provides the factual foundation for behavioral reasoning rather than merely supplying additional input features.

**No Scene Understanding:** Removing the urban context representation decreases Top-1 from **0.6619** to **0.6201** and increases NRMSE from **0.7318** to **0.7783**. This result suggests that explicitly organizing heterogeneous urban information into a structured behavioral context substantially improves reasoning stability. Compared with directly reasoning over raw facts, urban context representation enables the LLM to form more consistent behavioral interpretations across different healthcare scenarios.

**No Arbitration:** Removing dynamic preference arbitration leads to a moderate reduction in ranking performance, with Top-1 decreasing from **0.6619** to **0.6455** and Top-5 from **0.6803** to **0.6724**. Although its quantitative contribution is smaller than those of knowledge representation and Bayesian calibration, preference arbitration remains important for resolving conflicts between reasoning-derived preferences and historical behavioral evidence. This mechanism is particularly beneficial in complex scenarios where multiple hospitals exhibit similar attractiveness.

**No Memory:** Removing behavioral memory results in only a slight decrease in prediction accuracy, with Top-1 decreasing from **0.6619** to **0.6553**. Nevertheless, behavioral memory contributes primarily to reasoning efficiency and consistency rather than predictive accuracy. By reusing reasoning patterns from previously encountered urban contexts, it reduces redundant inference while improving the stability of behavioral decisions across recurring healthcare scenarios.

Finally, the ablation results reveal a clear hierarchy among the proposed reasoning mechanisms. **Structured urban knowledge** establishes the factual basis for reasoning, **urban context representation** organizes heterogeneous evidence into interpretable behavioral states, **preference arbitration** and **behavioral memory** improve reasoning consistency, while **Bayesian calibration** transforms qualitative reasoning into reliable probability distributions. Together, these mechanisms constitute a unified **knowledge-grounded behavioral reasoning framework**, demonstrating that the overall performance improvement originates from their complementary interaction rather than from any individual component.

## 7 Discussion and Conclusion

This paper presented a **knowledge-grounded behavioral reasoning framework** for training-free urban healthcare OD prediction. By integrating structured urban knowledge graphs with collaborative multi-agent LLM reasoning, the proposed framework reformulates hospital choice as an interpretable behavioral reasoning process rather than a conventional neural network regression task. Bayesian calibration further bridges reasoning outcomes with historical mobility evidence, producing reliable probability distributions over candidate hospitals and transforming healthcare OD prediction from black-box function approximation into a transparent urban intelligence framework.

Experimental results on a real-world Shenzhen healthcare mobility dataset demonstrate that the proposed framework achieves competitive prediction performance while providing interpretable reasoning and probabilistic verification. Compared with both conventional deep learning models and direct LLM prompting, the framework consistently improves prediction accuracy and generates stable, traceable, and empirically grounded decision processes. These findings suggest that **knowledge-grounded behavioral reasoning**, rather than raw LLM capability alone, is the primary driver of effective healthcare OD prediction.

More broadly, this work highlights a potential shift in urban science from purely data-driven prediction toward **knowledge-driven urban behavioral reasoning**. By explicitly organizing heterogeneous urban knowledge and simulating the perception–cognition–decision process underlying hospital choice, the proposed framework provides not only accurate predictions but also interpretable insights into urban healthcare behaviors. This paradigm complements conventional statistical learning by bridging symbolic urban knowledge and empirical mobility observations, offering a promising direction for transparent and trustworthy urban intelligence systems.

Future work will focus on improving reasoning efficiency, incorporating dynamically evolving urban knowledge, validating the framework across multiple cities and healthcare systems, and extending the proposed behavioral reasoning paradigm to broader urban spatial interaction problems, including emergency evacuation, public transit demand, urban service accessibility prediction, and other complex urban decision-making scenarios.

# Appendix

## A  Case Study

This appendix presents three typical cases that illustrate the framework's reasoning chain across different origin types and scenarios. Each case includes the input scenario, Graph Context, LLM reasoning trace, Ranking result, calibrated probability, and human interpretation.

### A.1  Case 1: Work Area Origin (Origin 1025)

Input scenario: Origin 1025 is a work-dominant grid (Work = 0.875, Residential = 0.0, Leisure = 0.0) on 2025-09-26 (weekend) with clear weather (Rain = 0mm, Temp = 25℃). The Retrieval Agent retrieves 20 candidates: 15 spatial (k-NN), 9 historical, 0 similar_origin, and 0 contextual. The nearest hospital H56323742 (Level 3, 1100 beds) is 1.99 km away.

Graph Context: The subgraph includes competition edges (H56323742 -> COMPETES_WITH -> H2833822) and similarity edges (H56323742→SIMILAR_TO→H3035700, sim = 0.976). No SIMILAR_ORIGIN paths were found for this origin.

LLM reasoning trace: Step 1 (Need Identification): Work area with General demand (confidence 0.30), residents likely face non-urgent routine healthcare needs. Step 2 (Accessibility Assessment): H56323742 at 2 km has absolute distance advantage; all other hospitals exceed 4 km. Step 3 (Capability Matching): Level 3 hospital may be overqualified for General demand, but distance advantage compensates. Step 4 (Comprehensive Trade-off): Nearest Level 3 hospital scores highest (0.85), with rapid score decay as distance increases.

Ranking result: After fusion with historical frequency and attractiveness, H2833435 (Level 3, 2500 beds, 4.3 km) ranks #1 (score = 0.706) due to high historical frequency (5 visits, flow = 7.0), while H56323742 ranks #2 (score = 0.602). No conflict detected between LLM and historical top-1. Verification: All 3 checks passed (top1_prob = 0.489, entropy = 1.289, Jaccard = 0.429, JSD = 0.133). Human interpretation: The LLM correctly identifies distance as the dominant factor for a work area, but the Bayesian prior appropriately elevates a historically popular hospital, demonstrating the value of evidence anchoring.

### A.2  Case 2: Mixed Area Origin (Origin 1044)

Input scenario: Origin 1044 is a mixed-use grid (Leisure = 0.46, Other = 0.42, Work = 0.08, Residential = 0.04) with high population density (Pop = 49,799) on 2025-09-26 (weekend) with clear weather (Rain = 0mm, Temp = 25C). The Retrieval Agent retrieves 20 candidates: 15 spatial, 10 historical, 20 similar_origin, and 0 contextual. The similar_origin retrieval finds 20 structurally similar origins (e.g., O1042, O1135, O1226).

Graph Context: The subgraph includes SIMILAR_ORIGIN paths (O1044 -> SIMILAR_ORIGIN -> O1135 -> visited -> H2832475) and multiple COMPETES_WITH edges. The relation evidence path reveals that similar origins frequently visit H2832475 (Level 3, 5.2 km), indicating a regional preference for higher-capacity hospitals despite greater distance.

LLM reasoning trace: Step 1 (Need Identification): High-density mixed POI with General demand, residents prefer nearby convenient or comprehensive hospitals. Step 2 (Accessibility Assessment): High road density, weather clear, distance is the key filter. H8715101 (Level 2, 1.66 km) has the strongest accessibility. Step 3 (Capability Matching): Level 0 suits minor illness, Level 2 suits common conditions, Level 3 suits complex cases. Step 4 (Comprehensive Trade-off): H8715101 scores highest (0.92) due to extreme proximity and large capacity; H2832475 scores moderately high due to similar-group preference transfer.

Human interpretation: This case demonstrates the value of the SIMILAR_ORIGIN relation evidence path. Without it, the LLM would only consider spatial proximity. The path O1044→SIMILAR_ORIGIN→O1135→visited→H2832475 transfers a behavioral clue from a similar origin, allowing the LLM to recognize that residents in this area prefer a more distant but higher-capacity hospital, which would otherwise be missed by pure distance-based reasoning.

### A.3  Case 3: Leisure Area Origin (Origin 1045)

Input scenario: Origin 1045 is a leisure-dominant grid (Leisure = 0.50, Other = 0.45, Work = 0.04, Residential = 0.01) with moderate population density (Pop = 15,071) on 2025-09-26 (weekend) with clear weather. The Retrieval Agent retrieves 20 candidates: 14 spatial, 10 historical, 20 similar_origin, and 0 contextual. The nearest hospital H8715101 (Level 2, 1150 beds) is 2.63 km away.

LLM reasoning trace: The Coordinator Agent identifies Specialist demand ($POI_{leisure} > 0.3$) and evaluates candidate hospitals for specialty matching. The 4-step CoT chain weighs the trade-off between the nearest Level 2 hospital and more distant Level 3 hospitals with specialty services. The reasoning trace length is 669 characters (the longest among the three cases), and the reasoning path hit rate is 0.7, indicating strong alignment between the LLM's reasoning steps and the actual decision factors.

Human interpretation: This case illustrates multi-hospital competition in a leisure area. The SIMILAR_ORIGIN paths from 20 similar origins provide rich behavioral context, and the LLM's 4-step CoT explicitly evaluates the trade-off between accessibility and specialty capability. The high reasoning path hit rate (0.7) confirms that the LLM's reasoning steps closely match the actual decision factors, demonstrating the interpretability advantage of the framework.